\RequirePackage{lineno}
\documentclass[aps,prl,showpacs,twocolumn,superscriptaddress,letterpaper]{revtex4}

\usepackage{graphicx}% Include figure files
\usepackage{dcolumn}% Align table columns on decimal point
\usepackage{bm}% bold math
\usepackage{amsmath, amssymb}% 
\usepackage{epstopdf}
\usepackage{color}
\usepackage{lipsum}
\usepackage{hhline}
\usepackage{multirow}
\usepackage{float}
\floatstyle{plaintop}
\restylefloat{table}
\usepackage[]{changes} % final
\begin{document}
%\linenumbers

\title{Measurement of Nuclear Transparency Ratios for Protons and Neutrons}

%%%%%%%%%%%%%%%%%%%%%%%%%%%%%%%%%%%%%%%%%%%% 
\newcommand*{\TAU }{School of Physics and Astronomy, Tel Aviv University, Tel Aviv 69978, Israel}
\newcommand*{\TAUindex}{1}
\affiliation{\TAU} 
\newcommand*{\MIT }{Massachusetts Institute of Technology, Cambridge, Massachusetts 02139, USA}
\newcommand*{\MITindex}{2}
\affiliation{\MIT} 
\newcommand*{\ODU}{Old Dominion University, Norfolk, Virginia 23529}
\newcommand*{\ODUindex}{3}
\affiliation{\ODU} 
\newcommand*{\UTFSM}{Universidad T\'{e}cnica Federico Santa Mar\'{i}a, Casilla 110-V Valpara\'{i}so, Chile}
\newcommand*{\UTFSMindex}{4}
\affiliation{\UTFSM}

\newcommand*{\ANL}{Argonne National Laboratory, Argonne, Illinois 60439}
\newcommand*{\ANLindex}{5}
\affiliation{\ANL}
\newcommand*{\ASU}{Arizona State University, Tempe, Arizona 85287-1504}
\newcommand*{\ASUindex}{6}
\affiliation{\ASU}
\newcommand*{\CMU}{Carnegie Mellon University, Pittsburgh, Pennsylvania 15213}
\newcommand*{\CMUindex}{7}
\affiliation{\CMU}
\newcommand*{\CUA}{Catholic University of America, Washington, D.C. 20064}
\newcommand*{\CUAindex}{8}
\affiliation{\CUA}
\newcommand*{\SACLAY}{IRFU, CEA, Universit'e Paris-Saclay, F-91191 Gif-sur-Yvette, France}
\newcommand*{\SACLAYindex}{9}
\affiliation{\SACLAY}
\newcommand*{\CNU}{Christopher Newport University, Newport News, Virginia 23606}
\newcommand*{\CNUindex}{10}
\affiliation{\CNU}
\newcommand*{\UCONN}{University of Connecticut, Storrs, Connecticut 06269}
\newcommand*{\UCONNindex}{11}
\affiliation{\UCONN}
\newcommand*{\DUKE}{Duke University, Durham, North Carolina 27708-0305}
\newcommand*{\DUKEindex}{12}
\affiliation{\DUKE}
\newcommand*{\FU}{Fairfield University, Fairfield CT 06824}
\newcommand*{\FUindex}{13}
\affiliation{\FU}
\newcommand*{\FERRARAU}{Universita' di Ferrara , 44121 Ferrara, Italy}
\newcommand*{\FERRARAUindex}{14}
\affiliation{\FERRARAU}
\newcommand*{\FIU}{Florida International University, Miami, Florida 33199}
\newcommand*{\FIUindex}{15}
\affiliation{\FIU}
\newcommand*{\FSU}{Florida State University, Tallahassee, Florida 32306}
\newcommand*{\FSUindex}{16}
\affiliation{\FSU}
\newcommand*{\Genova}{Universit$\grave{a}$ di Genova, 16146 Genova, Italy}
\newcommand*{\Genovaindex}{17}
\affiliation{\Genova}
\newcommand*{\GWUI}{The George Washington University, Washington, DC 20052}
\newcommand*{\GWUIindex}{18}
\affiliation{\GWUI}
\newcommand*{\ISU}{Idaho State University, Pocatello, Idaho 83209}
\newcommand*{\ISUindex}{19}
\affiliation{\ISU}
\newcommand*{\INFNFE}{INFN, Sezione di Ferrara, 44100 Ferrara, Italy}
\newcommand*{\INFNFEindex}{20}
\affiliation{\INFNFE}
\newcommand*{\INFNFR}{INFN, Laboratori Nazionali di Frascati, 00044 Frascati, Italy}
\newcommand*{\INFNFRindex}{21}
\affiliation{\INFNFR}
\newcommand*{\INFNGE}{INFN, Sezione di Genova, 16146 Genova, Italy}
\newcommand*{\INFNGEindex}{22}
\affiliation{\INFNGE}
\newcommand*{\INFNRO}{INFN, Sezione di Roma Tor Vergata, 00133 Rome, Italy}
\newcommand*{\INFNROindex}{23}
\affiliation{\INFNRO}
\newcommand*{\INFNTUR}{INFN, Sezione di Torino, 10125 Torino, Italy}
\newcommand*{\INFNTURindex}{24}
\affiliation{\INFNTUR}
\newcommand*{\ORSAY}{Institut de Physique Nucl'eaire, IN2P3-CNRS, Universit'e Paris-Sud, Universit'e Paris-Saclay, F-91406 Orsay, France}
\newcommand*{\ORSAYindex}{25}
\affiliation{\ORSAY}
\newcommand*{\ITEP}{Institute of Theoretical and Experimental Physics, Moscow, 117259, Russia}
\newcommand*{\ITEPindex}{26}
\affiliation{\ITEP}
\newcommand*{\JMU}{James Madison University, Harrisonburg, Virginia 22807}
\newcommand*{\JMUindex}{27}
\affiliation{\JMU}
\newcommand*{\KNU}{Kyungpook National University, Daegu 41566, Republic of Korea}
\newcommand*{\KNUindex}{28}
\affiliation{\KNU}
\newcommand*{\LAMAR}{Lamar University, 4400 MLK Blvd, PO Box 10009, Beaumont, Texas 77710}
\newcommand*{\LAMARindex}{29}
\affiliation{\LAMAR}
\newcommand*{\MISS}{Mississippi State University, Mississippi State, MS 39762-5167}
\newcommand*{\MISSindex}{30}
\affiliation{\MISS}
\newcommand*{\UNH}{University of New Hampshire, Durham, New Hampshire 03824-3568}
\newcommand*{\UNHindex}{31}
\affiliation{\UNH}
\newcommand*{\NSU}{Norfolk State University, Norfolk, Virginia 23504}
\newcommand*{\NSUindex}{32}
\affiliation{\NSU}
\newcommand*{\OHIOU}{Ohio University, Athens, Ohio  45701}
\newcommand*{\OHIOUindex}{33}
\affiliation{\OHIOU}
\newcommand*{\RPI}{Rensselaer Polytechnic Institute, Troy, New York 12180-3590}
\newcommand*{\RPIindex}{34}
\affiliation{\RPI}
\newcommand*{\URICH}{University of Richmond, Richmond, Virginia 23173}
\newcommand*{\URICHindex}{35}
\affiliation{\URICH}
\newcommand*{\ROMAII}{Universita' di Roma Tor Vergata, 00133 Rome Italy}
\newcommand*{\ROMAIIindex}{36}
\affiliation{\ROMAII}
\newcommand*{\MSU}{Skobeltsyn Institute of Nuclear Physics, Lomonosov Moscow State University, 119234 Moscow, Russia}
\newcommand*{\MSUindex}{37}
\affiliation{\MSU}
\newcommand*{\SCAROLINA}{University of South Carolina, Columbia, South Carolina 29208}
\newcommand*{\SCAROLINAindex}{38}
\affiliation{\SCAROLINA}
\newcommand*{\TEMPLE}{Temple University,  Philadelphia, PA 19122 }
\newcommand*{\TEMPLEindex}{39}
\affiliation{\TEMPLE}
\newcommand*{\JLAB}{Thomas Jefferson National Accelerator Facility, Newport News, Virginia 23606}
\newcommand*{\JLABindex}{40}
\affiliation{\JLAB}
\newcommand*{\GLASGOW}{University of Glasgow, Glasgow G12 8QQ, United Kingdom}
\newcommand*{\GLASGOWindex}{41}
\affiliation{\GLASGOW}
\newcommand*{\YORK}{University of York, York YO10, United Kingdom}
\newcommand*{\YORKindex}{42}
\affiliation{\YORK}
\newcommand*{\VIRGINIA}{University of Virginia, Charlottesville, Virginia 22901}
\newcommand*{\VIRGINIAindex}{43}
\affiliation{\VIRGINIA}
\newcommand*{\WM}{College of William and Mary, Williamsburg, Virginia 23187-8795}
\newcommand*{\WMindex}{44}
\affiliation{\WM}
\newcommand*{\YEREVAN}{Yerevan Physics Institute, 375036 Yerevan, Armenia}
\newcommand*{\YEREVANindex}{45}
\affiliation{\YEREVAN}
 \newcommand*{\CSUDH}{California State University Dominguez Hills, Carson, CA 90747}
\newcommand*{\CSUDHindex}{46}
\affiliation{\CSUDH}
\newcommand*{\NOWISU}{Idaho State University, Pocatello, Idaho 83209}
\newcommand*{\NOWINFNGE}{INFN, Sezione di Genova, 16146 Genova, Italy}
\newcommand*{\NRCN}{Nuclear Research Centre Negev, Beer-Sheva, Israel}
\newcommand*{\NRCNindex}{47}

%%%%%%%%%%%%%%%%%%%% authors %%%%%%%%% 
\author{M.~Duer}
\affiliation{\TAU}
\author{O.~Hen}
\email[Contact Author \ ]{hen@mit.edu}
\affiliation{\MIT}
\author{E.~Piasetzky}
\affiliation{\TAU}
\author{L.B.~Weinstein}
\affiliation{\ODU}
\author{A.~Schmidt}
\affiliation{\MIT}
\author{I.~Korover}
\affiliation{\TAU}
\author{E. O. Cohen}
\affiliation{\TAU}
\author {H.~Hakobyan} 
\affiliation{\UTFSM}
\author {S. Adhikari}
\affiliation{\FIU}
\author {G. Angelini} 
\affiliation{\GWUI}
% ...
\collaboration{The CLAS Collaboration}
\noaffiliation

\begin{abstract}
%Nucleon knockout reaction measurements are an invaluable guide to understanding nuclear and nucleon structure. 
%Knowledge of nuclear transparency in such reactions serves as a baseline for the identification of in-medium, 
%bound-nucleon modifications, color-transparency, and reaction rate estimations, as well as a broad range of 
%phenomena in high-energy, heavy-ion, and hadronic physics. The experimental study of nuclear transparency is 
%therefore crucial for the theoretical description and interpretation of experimental data. 
This paper presents, for the first time, measurements of neutron transparency ratios for nuclei relative to C measured using the 
$(e,e'n)$ reaction, spanning measured neutron momenta of 1.4 to 2.4 GeV/$c$. The transparency ratios were 
extracted in two kinematical regions, corresponding to knockout of mean-field nucleons and to the breakup of 
Short-Range Correlated nucleon pairs. The extracted neutron transparency ratios are consistent with each other 
for the two measured kinematical regions and agree
%, within experimental uncertainties, 
with the proton transparencies extracted from new and 
previous $(e,e'p)$ measurements, including those from neutron-rich nuclei such as lead. The data also agree
with and confirm the Glauber approximation that is commonly used to interpret experimental data. The 
nuclear-mass-dependence of the extracted transparencies scales as $A^\alpha$ with $\alpha = -0.289 \pm 0.007$, 
which is consistent with nuclear-surface dominance of the reactions.

\end{abstract}

\maketitle

% =====================================================================

%introduction
%High-energy, large-momentum-transfer electron scattering reactions are
%used to study a wide range of nuclear phenomena, ranging from the
%shell-structure of nuclei to quark-gluon color transparency effects of the nucleon
%\cite{,,,,,,Dutta12}.

High-energy, large-momentum-transfer electron scattering reactions are used 
to study a wide range of nuclear phenomena, including the shell-structure of 
nuclei~\cite{Hofstadter:1956qs,DeForest:1966ycn,kelly96}, the modification of 
the quark-gluon substructure of protons and neutrons by the nuclear medium~\cite{Frankfurt88,Geesaman95,Norton03,Weinstein:2010rt, Hen:2013oha,Barak:Nature2019}, 
the properties of nucleon correlations~\cite{subedi08,korover14,Atti:2015eda,Hen:2016kwk,Meytal}, and quark-gluon color transparency 
effects~\cite{Frankfurt:1992dx,Makins:1994mm,Frankfurt2,Sargsian02,Dutta12}. 
In many experiments, information about the nucleus is gleaned from 
the detection of knocked out nucleons, or other hadrons produced in the 
reaction, after they exit the nucleus. The quantitative interpretation of such
experiments relies on modeling both the electron interaction cross-section and
the propagation of hadrons through the atomic nucleus. The nuclear transparency 
factor, $T(A)$, describes the probability of an outgoing hadron to emerge from 
the nucleus, quantifying the multiple scattering of the hadron with the 
surrounding nucleons~\cite{garino92,Makins:1994mm,oneill95,abbott98,garrow02,hen12a}.

At high-energies, nuclear transparency can be calculated using the Glauber approximation 
\cite{Frankfurt:1996xx,Ryckebusch:2003fc,panda,Wim_cx,Miller:2007ri,Benhar:1991dd}, 
and such calculations serve as a key ingredient in the analyses of nuclear~\cite{Frankfurt2,Duer:2018sxh}, 
hadronic~\cite{Clasie:2007aa,alvioli10,Dutta12}, neutrino~\cite{Martinez:2005xe,Benhar:2005dj}, 
and heavy-ion~\cite{PhysRevC.88.044909,Broniowski:2007ft,Charagi:1990zz,Alver:2008aq,Frankfurt:2002sv}, 
physics experiments.
Thus, the experimental extraction of transparency factors for single-nucleon knockout 
reactions at different kinematics and for different nuclei serves both as an 
important test of the Glauber approximation and as a baseline for obtaining 
information on the structure and dynamics of individual nucleons bound in 
nuclei. See~\cite{Sargsian02,Miller:2007ri,Dutta12} for recent reviews.

We present an experimental extraction of the transparency ratios,
$T(A) / T(^{12}$C$)$, for proton and neutron knockout, from
measurements of Quasi-Elastic (QE) $A(e,e'n)$ and $A(e,e'p)$
reactions. Here $A$ represents carbon, aluminum, iron, and lead (or
their corresponding atomic numbers). The carbon nucleus was chosen as
a reference because it is a well-studied symmetric nucleus. The
transparency ratios were extracted in two kinematic regimes, the first
corresponding to the knockout of Mean-Field (MF) nucleons and the
second corresponding to the knockout of nucleons from Short-Range
Correlated (SRC) nucleon pairs~\cite{Hen:2016kwk}. In both regimes, a
requirement of large momentum transfer ($Q^2 \ge 1.5$ GeV$^2$)
was enforced, which constrained the detected nucleon momentum to be
greater than $1.4$ GeV/c.

In MF kinematics, the reconstructed initial momenta of the knockout
nucleons are below the nuclear Fermi momentum, where the nuclear
spectral functions are relatively well-modeled. In SRC kinematics, the
reconstructed initial momenta of the knockout nucleons are greater
than the nuclear Fermi momentum because the nucleons are members of
SRC pairs~\cite{Hen:2016kwk}.

The extracted transparencies for the two kinematics are consistent
with each other within their experimental uncertainties. The neutron
knockout transparencies, obtained here for the first time, are also
consistent with the proton knockout transparencies, even in nuclei
such as lead, which has about 1.5 times more neutrons than
protons. The $A$-dependences of the neutron- and proton-knockout
transparencies show a power-law scaling of $A^\alpha$,
%(where $A$ is the nuclear mass number), with a scale factor 
with $\alpha = -0.289 \pm 0.007$. This exponent is consistent  
with $-1/3$, the value expected for scattering from the nuclear surface.  
Moreover, the proton knockout data agree with Glauber calculations, 
validating their use in analyses of high-energy nuclear reactions.  
While Glauber calculations of neutron knockout reactions are unavailable 
at the moment, they should not significantly deviate from the proton 
knockout calculations in order to be consistent with the data presented here.

The analysis presented here was carried out as part of the Thomas
Jefferson National Accelerator Facility (JLab) Hall B Data-Mining
project~\cite{DataMining2}.
% using data from the Experiments E02-104 and
%E02-110, which took place in 2004 in Hall~B at JLab. 
The experiment used a 5.01 GeV unpolarized electron beam incident on a dedicated
target system \cite{Hakobyan:2008zz}.
%that held both a nuclear target and a deuterium cryogenic cell simultaneously in the beamline. 
Scattered electrons and knockout nucleons were detected using the
CEBAF Large Acceptance Spectrometer (CLAS)~\cite{Mecking:2003zu}.

CLAS consisted of a toroidal super-conducting magnet and six
independent sectors containing sets of drift chambers (DC),
time-of-flight scintillation counters (TOF), Cherenkov counters (CC),
and electro-magnetic calorimeters (EC) for trajectory reconstruction
and particle identification. The polar angular acceptance was $8^\circ
< \theta < 140^\circ$. The azimuthal angular acceptance was 50\% at
small polar angles, increasing to 80\% at larger polar angles.

The specially designed target consisted of a 2-cm liquid deuterium (LD$_2$)
cryotarget, followed by one of five alternately-installable solid
targets ranging in thickness from 0.16 g/cm$^2$ to 0.32 g/cm$^2$ (thin
and thick Al, C, Fe, and Pb, all in natural isotopic abundances)
\cite{Hakobyan:2008zz}. 
The experiment collected most of its integrated luminosity on the C and Fe
targets, with reduced statistics on Al and Pb.
The LD$_2$ target cell and the solid targets were separated by about 4 cm. 
The beam lost an average of 0.6 MeV energy in the LD$_2$ target with negligible contribution to its radiative tail~\cite{Hakobyan:2008zz}.
We selected events with particles
scattering from the solid target by reconstructing the intersections
of their trajectories with the beam line. The vertex reconstruction
resolution for both electrons and protons was sufficient to
unambiguously separate particles originating from the cryotarget and
the solid target. The analysis presented here used only data from the
solid targets.

Scattered electrons were identified and pions rejected by requiring
negatively-charged tracks that produced more than 2.5 photo-electrons
in the CC with energy deposits in the inner and outer parts of the EC
that are correlated and proportional to the particle momentum
~\cite{Mecking:2003zu}. Neutrons were identified by observing
interactions in the forward EC (covering polar angles from 8$^\circ$ to
45$^\circ$), with no associated hit in the preceding TOF bar and
no charged-particle tracks in the DC. 
We only considered interactions
that were within the fiducial region of the EC, defined to include
only hits reconstructed at least 10 cm away from its edge. 
Due to the lack of sufficient Hydrogen data, the angle- and
momentum-dependent neutron detection efficiency and momentum
reconstruction resolution were determined simultaneously using the
exclusive $d(e,e'p\pi^+\pi^-)n$ and $d(e,e'p\pi^+\pi^-n)$ reactions
\cite{TAU_Theis,Meytal} that, for our beam energy, have higher statistics and preferable 
kinematical coverage than the commonly used  $d(e,e'p)n$ and $d(e,e'pn)$ reactions.
As both the nuclear and deuterium targets were in the beam simultaneously, the random-coincidence background was the same for the efficiency and physics measurements. Due to the low luminosity of the measurement such backgrounds were small and do not contribute to the systematic uncertainty.
Protons in CLAS were identified by requiring
that the difference between the measured time of flight and that
calculated from the measured momentum and the nominal proton mass 
be within two standard deviations from the mean of the event distribution 
over that difference. 
This cut is an optimized trade-off between excellent separation and minimal 
Monte Carlo correction for lost events and clearly separates protons from pions 
and kaons up to a momentum of $2.8$ GeV/$c$.

% ----------------------------------------------------------------
%% * = Not a standard figure, as we want to span it on the two columns
\begin{figure*}%[t]
\centering
\includegraphics[width=14 cm]{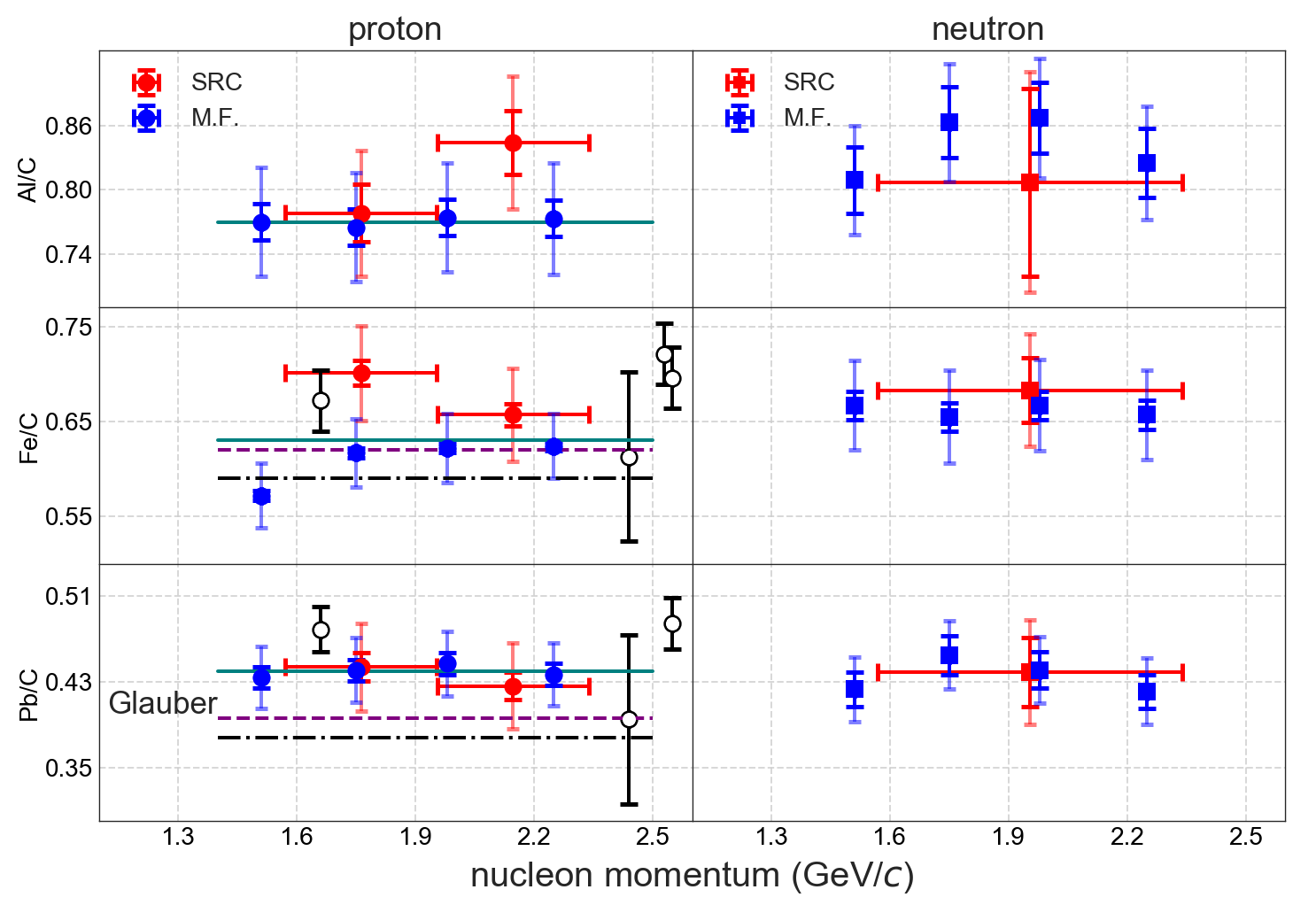} %width=\columnwidth, height=7.5cm
\caption{\label{fig:0} The estimated transparency ratios for MF and SRC kinematics,
both for protons and neutrons, as a function of the nucleon momentum.
Inner error bars are statistical and outer error bars include statistical and systematics uncertainties, the latter are common for the different data-points of a given measurement.
The black open circles show the world data for the transparency ratios for MF proton knockout from Ref.~\cite{Oniell,Garino,abbott98,garrow02}. Glauber calculations are shown as dot-dashed~\cite{Frankfurt2}, dashed~\cite{panda}, and solid~\cite{Wim_cx,WimPrivate} lines. The nucleon momentum range for the SRC data points is denoted by the horizontal line round each point, while that of the MF data points is the same for all points and is not shown for clarity.}
\end{figure*}
% ----------------------------------------------------------------

We applied fiducial cuts on the angles and momenta of all charged
particles to avoid regions with steeply varying acceptance close to
the magnetic coils of CLAS and the edges of the detectors. The
fiducial region for neutron detection was defined by the EC fiducial
region as  described above. In order to match the knockout nucleon
acceptances in the $A(e,e'p)$ and the $A(e,e'n)$ reactions, we
considered only protons or neutrons emerging from the nucleus at angles that 
are within the fiducial regions for both neutrons and protons.  Also,
we considered only neutrons or 
protons with momenta up to 2.4 GeV/$c$, the maximum
neutron momentum considered in the analysis~\cite{TAU_Theis,Meytal}.

%We corrected the kinetic energy of the incoming electron and of the
%scattered electron and proton for Coulomb distortions on an event-by-event basis using the Effective Momentum Approximation (EMA)~\cite{Aste05}. %Following~\cite{Fomin:2012}, we assumed an effective electric potential equal to 75\% of the potential produced by $Z$ unscreened charges located at the %center of the nucleus. This amounts to a $3$-, $5$-, $10$- and $20$-MeV correction for C, Al, Fe, and Pb, respectively~\cite{hen12a,hen14}.

We corrected the kinetic energy of the incoming electron and of the scattered electron and proton for Coulomb distortions on an event-by-event basis using the Effective Momentum Approximation (EMA)~\cite{Aste05}, as detailed in Ref.~\cite{hen12a,hen14}.

In QE electron-scattering processes, within the one-body, one-photon
exchange approximation, the momentum and the energy transfer to the
virtual photon are absorbed by a single nucleon with an initial momentum $\vec{p}_i$,
leaving the undetected (missing) residual ($A-1$) system with
excitation energy $E_{miss} = \omega - T_N -T_{A-1}$, where $\omega$
is the virtual-photon energy transfer, $T_{N}$ is the kinetic energy
of the detected nucleon, and $T_{A-1}$ is the reconstructed kinetic
energy of the residual $A-1$ system. The missing momentum is 
${p}_{miss} = |\vec{p}_N - \vec{q}|$, where
$\vec{p}_N$ is the measured nucleon momentum and $\vec{q}$ is the
momentum of the virtual photon. Using this definition, in the Plane
Wave Impulse Approximation (PWIA), the initial momentum of the struck
nucleon ${p}_{i}$ is equal to ${p}_{miss}$.

The identification of $A(e,e'N)$ events (in which $N$ stands for
either proton or neutron) in either MF or SRC kinematics, follows
previous work~\cite{hen14,hen12a,Meytal}. 
%We required $Q^2 > 1.5$ GeV$^2$/$c^2$ for both kinematics.  
For the MF kinematics, this
includes selection of events with $Q^2 > 1.5$ GeV$^2$, and low
missing energy and momentum: $E_{miss} < 80 - 90$ MeV and $p_{miss} <
250$ MeV/$c$ \cite{Geesaman,Oniell,Garino} and $-0.05<y<0.25$. The
latter is a scaling variable that is related to the component of the
initial momentum of the knocked-out nucleon in the direction of the
momentum transfer vector~\cite{Donnelly:2002fca}. For QE events $y$ is
expected to be centered around zero. In this analysis it is shifted
to small positive values due to the limited angular acceptance of the
EC. For the SRC kinematics, this includes selection of events with
$x_B > 1.2$, and a
nucleon with momentum magnitude in the range $0.62 <
|\vec{p}_N| / |\vec{q}| < 0.96$ and direction similar to that of the
virtual photon ($\theta_{Nq} < 25^\circ$). Also, in SRC kinematics, we
required that the missing momentum was high (${p}_{miss} > 300$
MeV/$c$)~\cite{hen12a}.

Although these event-selection cuts proved effective for studies of $(e,e'p)$ reactions in the two relevant kinematics, in the case of $(e,e'n)$ the neutron momentum-reconstruction resolution was insufficient to directly distinguish between events with high and low missing momentum and energy. To overcome this issue, we developed an alternative set of cuts that makes use only of quantities insensitive to the poor neutron momentum-reconstruction resolution, including the momentum of the scattered electron and the direction of the knockout nucleon. This alternative set of cuts was chosen to select an event sample as similar as possible to the nominal set, with reduced sensitivity to the neutron momentum reconstruction resolution~\cite{Meytal}.

The development of these alternative selection cuts required analyzing a data-driven Monte-Carlo $(e,e'p)$ event sample, in which each measured event was used to generate several events for which the measured momentum of the proton was smeared according to the neutron momentum resolution. For each generated event, all related kinematical variables, such as missing momentum and energy, were calculated twice, using the proton measured and smeared momentum. This allowed us to estimate the efficiency and purity of the alternative selection cuts when applied to $(e,e'n)$ events.
%(that used the smeared kinematical variables) while keeping track of the purity of the selected event sample as compared to the original sample (obtained with the nominal MF and SRC selection cuts described above).

The alternative selection cuts are listed in Table I for both MF and SRC kinematics.
In both cases, the alternative cuts
were chosen such that the selection efficiency is as large as
possible while background contamination by other events is
minimal. The efficiency of the alternative cuts is about $90\%$ in
either kinematics, with a contamination of about $10\%$, making the
total event sample sizes equivalent to those obtained using the
unsmeared momentum and the nominal selection cuts (MF and SRC).

See online supplementary materials for additional details on the analysis, 
including comparisons between the kinematical distributions 
of the selected $(e,e'n)$ and smeared $(e,e'p)$ event samples.

Using the selected event samples, we then extracted the cross-section
ratios for scattering off the solid targets relative to carbon. The
cross-section ratios are defined as the normalized-yield ratios,
where each yield is normalized to the number of scattering centers
(number of protons or number of neutrons) in the target and the total
number of electrons incident on the corresponding target during the
experiment. Detector acceptance effects cancel in the cross-section
ratios because (a) all solid targets were located at the same position
on the beam line and (b) the shapes of the event distributions giving 
the yields in the nominator and the denominator of each transparency ratio, 
were observed to be similar for  the final-state kinematical variables for the 
different targets (including missing energy and momentum).

The $A(e,e'N)$ cross-section ratios were corrected for radiative
effects~\cite{Radiation} in the same way as was done
in~\cite{hen12a,egiyan03,Egiyan06}. The radiative correction to the
transparency ratio was found to be $\sim1\%$, $5\%$, and 6\% for Al/C, Fe/C, and Pb/C ratio, respectively, with a negligible contribution to the corresponding total systematic uncertainty.

Nuclear transparency is formally defined as the ratio of the experimentally extracted nucleon knock-out cross-section to the Plane Wave Impulse Approximation (PWIA) cross-section,
\begin{equation}
T_N(A) = \frac{\sigma_{exp}A(e,e'N) }{ \sigma_{PWIA}A(e,e'N)}.
\label{eq:T}
\end{equation}
In the commonly used factorized approximation for large-$Q^2$ reactions, $\sigma_{PWIA}$ is given by (see~\cite{FOREST}):
\begin{equation}
\sigma_{PWIA}(e,e'N) = \frac{K}{\#N_{tar}} \cdot \sigma_{eN} \cdot \oint S_A(E,p_i)dEd^{3}p_{i},
\label{eq:PWIA}
\end{equation}
in which \#$N_{tar}$ is the number of relevant nucleons in the target nucleus (i.e. number of protons for $(e,e'p)$ and neutrons for $(e,e'n)$ reactions), $K = |\vec{p}_N| \cdot E_{N}$ is a kinematical factor, $E_{N}$ is the energy of the outgoing nucleon, $\sigma_{eN}$ is the off-shell
electron-nucleon elementary cross section, $S_{A}(E,p_i)$ is the
nuclear spectral function, which defines the probability for finding a
nucleon in the nucleus with momentum $p_i$ and separation energy
$E$. $S_{A}(E,p_i)$ is normalized as $\int_{0}^{\infty} S_{A}(E,p_i)
dE d^3p_i \equiv \#N_{tar}$. 
The limits of the integral over the spectral function in Eq.~\ref{eq:PWIA}
should correspond to the experimental acceptance.

Under the condition that the measurements for a nucleus with $A$ nucleons 
and for $^{12}$C are done in equivalent kinematics, as was done in this work, 
their transparency ratio is given by:
\begin{equation}
\frac{T_N(A)}{T_N(C)} = \frac{\sigma_{exp}A(e,e'N) }{ \sigma_{exp}C(e,e'N) } \cdot \frac{\oint S_C(E,p_i)dEd^{3}p_{i}}{\oint S_A(E,p_i)dEd^{3}p_{i}},
\label{eq:Tratio}
\end{equation}
in which the spectral functions for $A$ and $C$ are integrated over the same kinematical regions.

\begin{figure} [t]
\includegraphics[width=8cm, height=6.5cm]{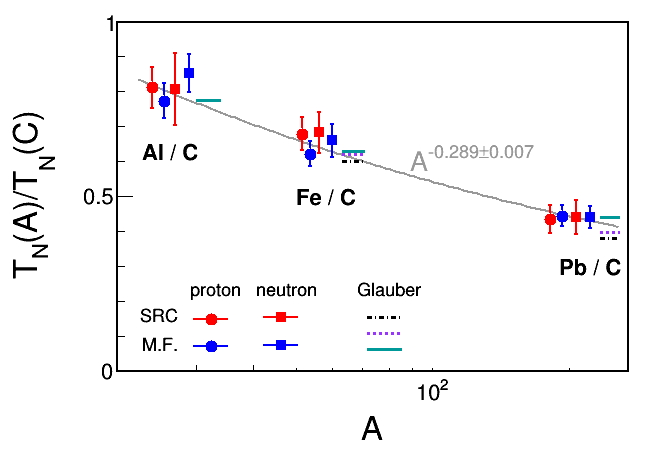}
\caption{\label{fig:0} The estimated transparency ratios for MF and SRC kinematics, both for protons and neutrons, together with a power law fit to a weighted average (grey line), as described in the text. For Fe and Pb nuclei, also shown are results based on three Glauber Calculations:~\cite{Frankfurt2} dotted line,~\cite{panda} dashed line, and~\cite{Wim_cx,WimPrivate} solid line.}
\end{figure}

For the MF kinematics, Eq.~\ref{eq:Tratio} can be expressed as:
\begin{equation}
\frac{T^{MF}_N(A)}{T^{MF}_N(C)} = \frac{\int_{0}^{k_0} n_C(p_i)dp_i}{\int_{0}^{k_0} n_A(p_i)dp_i} \cdot \frac{\sigma_{exp}A(e,e'N) }{ \sigma_{exp}C(e,e'N)},
\label{eq:Tratio_MF}
\end{equation}
where $\sigma_{exp}A(e,e'N) / \sigma_{exp}C(e,e'N)$ is the measured
nucleon knockout cross-section ratio discussed above and the first
term is the ratio of integrals over the mean-field part of the nuclear
momentum density, which, due to the large missing-energy cut, replaces
the integrals over the mean-field spectral functions. The nuclear
momentum density is defined as $n_A(p_i) \equiv \int_{0}^{\infty}
S_{A}(E,p_i) dE$. The later was calculated following~\cite{Frankfurt2}. The
integral calculations in Eq.~\ref{eq:Tratio_MF} were done using three
different models for the mean-field momentum distribution: Ciofi and
Simula~\cite{cda96}, Woods-Saxon~\cite{vanhalst12}, and
Serot-Walecka~\cite{Furnstahl} with $k_0$, the upper limit of the MF
momentum range, chosen to be the average between 300 MeV/$c$ and the
Fermi sea level, $k_F$ = 221; 260; 260; and 260 (280) MeV/$c$ for C,
Al, Fe, Pb, respectively, for protons (neutrons)~\cite{moniz71}. We
assigned the half difference between the two extreme values obtained
by considering the different values of $k_0$ and the different models
as a corresponding systematic uncertainty. The values of the latter
are 4.9\% (3.8\%), 4.2\% (5.7\%), and 4.3\% (4.5\%) for protons
(neutrons) for the Al/C, Fe/C, and Pb/C ratios, respectively. The
results of this calculation are consistent with those previously
obtained by Hartree-Fock-Slater wave functions~\cite{Frankfurt2}.

The transparency ratios in SRC kinematics are extracted following~\cite{hen12a} as:
\begin{equation}
\frac{T^{SRC}_N(A)}{T^{SRC}_N(C)} = \frac{1}{a_2(A/{\rm C})} \cdot \frac{\sigma_{exp}A(e,e'N) / A }{ \sigma_{exp}C(e,e'N) / 12},
\label{eq:Tratio_SRC}
\end{equation}
where $a_2(A/C)$ is the relative number of $2N$-SRC pairs per nucleon in nuclei $A$ and C. These ratios were adapted from~\cite{Barak} and are based on a compilation of world data for the $(e,e')$ cross-section ratio at large $Q^2$ and $x_B > 1$ with different theoretical corrections. 

\begin{table}[t]
\center
\caption{The $(e,e'N)$ event selection cuts for the MF kinematics and SRC kinematics.}
\begin{tabular}{|c|c|}
\hline
MF  & SRC  \\ \hline \hline
$-0.05<y<0.25$                   & $x_{B}>1.1$       \\ \hline
$0.95<\omega<1.7$ GeV    & $0.62<|\vec{p}_N|/|\vec{q}|<1.1$       \\ \hline
$\theta_{Nq}<8^{\circ}$       & $\theta_{Nq}<25^{\circ}$       \\ \hline
$p_{miss}<0.3$ GeV/$c$    & $m_{miss}<1.175$ GeV/$c^{2}$       \\ \hline
$E_{miss}<0.19$ GeV          & $0.4<p_{miss}<1$ GeV/$c$       \\ \hline
\end{tabular}
\end{table}

As shown in Fig.  1, the extracted transparency ratios are independent
of nucleon momentum between $1.4$ and $2.4$ GeV/$c$ for both protons
and neutrons, and for each of the three nuclei. Also shown are previous
measurements for protons~\cite{Oniell,Garino,abbott98,garrow02},
which are consistent with the new results. 
The proton knockout data also show an overall good agreemet with various 
Glauber calculations~\cite{Frankfurt2,panda,Wim_cx,WimPrivate}.

The transparency ratios,
averaged over nucleon momentum and type of nucleon, for each
kinematics, are listed in Tables II. The systematic uncertainties of
these estimates include the sensitivity to the
event selection cuts (see online supplementary materials Tables I and III), and a $2\%$ uncertainty of
the integrated charge. For the SRC the uncertainties also include the
uncertainty of $a_2(A/C)$ ($5\%$), and a $5\%$ uncertainty due to the
$np$-dominance assumption for the Pb/C case (see
Ref.~\cite{hen12a}). For the MF, the uncertainties also include the uncertainty from the MF integrals discussed above. The systematic uncertainty is independent of nucleon momentum. In Fig. 1 and elsewhere in this paper, the uncertainties shown and quoted are total uncertainties (systematic and statistical summed in quadrature), except if specifically stated otherwise. 

Figure 2 shows the extracted transparency ratios, averaged over the momentum
range shown in Fig. 1, as a function of the nuclear mass number. 
Notice that the momentum ranges are 1.64--2.34 GeV/$c$ (3
bins) for MF, and 1.57--2.34 GeV/$c$ for SRC. As mentioned above, the
transparency ratios are independent of nucleon momentum in these
ranges. For example, averaging the MF transparency ratios over
1.40--2.34 GeV/$c$ (all 4 bins) yields a value that is within 1\%
(much smaller than the smallest total uncertainty) of the average over
1.64--2.34 GeV/$c$ (last 3 bins).
The results are not sensitive to the smearing, since the smeared and un-smeared 
$(e,e'p)$ transparency ratios in each kinematics are the same (see online
supplementary materials).

Since all the four transparency
ratios are consistent with each other within their experimental
uncertainties, we take their weighted average for each nucleus
and, following \cite{Oniell,Garino,abbott98,garrow02}, we fit them to a
power law in the form of $A^{\alpha}$.
The Glauber calculation indicates the distribution of the hard process in the nucleus.
Our extracted value of $\alpha = -0.289 \pm 0.007$ is consistent with the Glauber result of $\alpha = -0.288$ to $-0.337$~\cite{Frankfurt2,panda,Wim_cx,WimPrivate}.
While the electron-nucleon interaction can take place anywhere in the nucleus, the observed power law indicates that the measured events  come predominantly from electron-nucleon interactions at the surface of the nucleus in all measured reactions, and also in neutron-rich nuclei like lead~\cite{Frankfurt2,panda,Wim_cx,WimPrivate}. Non-surface interactions are more susceptible to substantial nucleon-nucleon rescattering which removes the event from the experimental phase-space.

\begin{table}[t]
\center
\caption{The transparency ratios and their uncertainties for MF and SRC kinematics. The first uncertainty is
statistical and the second is systematical.}
\begin{tabular}{|c| c | c | c | c |}
\hline
\multicolumn{3}{|c|}{MF} \\
\hline
& $(e,e'p)$ & $(e,e'n)$ \\
\hline 
\footnotesize{Al/C} & \footnotesize {0.771$\pm$0.017$\pm$0.048} & \footnotesize {0.853$\pm$0.033$\pm$0.043} \\
\footnotesize{Fe/C} & \footnotesize {0.621$\pm$0.005$\pm$0.035} & \footnotesize {0.660$\pm$0.015$\pm$0.044} \\
\footnotesize{Pb/C} & \footnotesize {0.442$\pm$0.010$\pm$0.027} & \footnotesize {0.439$\pm$0.017$\pm$0.026} \\
\hline \hline
\multicolumn{3}{|c|}{SRC} \\
\hline 
\footnotesize{Al/C} & \footnotesize {0.811$\pm$0.028$\pm$0.053} & \footnotesize {0.807$\pm$0.088$\pm$0.053}\\
\footnotesize{Fe/C} & \footnotesize {0.679$\pm$0.013$\pm$0.046} & \footnotesize {0.683$\pm$0.034$\pm$0.048}\\
\footnotesize{Pb/C} & \footnotesize {0.435$\pm$0.013$\pm$0.038} & \footnotesize {0.439$\pm$0.032$\pm$0.037}\\
\hline 
\end{tabular}
\end{table}

In summary, we determined experimentally $A(e,e'p)$ and $A(e,e'n)$
cross-section ratios for $^{27}$Al, $^{56}$Fe, and $^{208}$Pb nuclei
relative to $^{12}$C in MF and SRC kinematics. From these ratios we
extracted the nuclear transparency ratios for protons and neutrons in
each kinematics. Both the proton and neutron transparency ratios are
independent of nucleon momentum and  consistent with each other
within the experimental uncertainty, for all measrued nuclei. The
$A$-dependence of the transparency ratio as well as the Glauber
calculations are consistent with the hypothesis that nucleon knockout
occurs on the nuclear surface.

% ====================================================================

\begin{acknowledgments}
We acknowledge the efforts of the staff of the Accelerator and Physics
Divisions at Jefferson Lab that made this experiment possible. We are
also grateful for many fruitful discussions with L.L. Frankfurt,
M. Strikman, J. Ryckebusch, W. Cosyn, M. Sargsyan, and C. Ciofi degli
Atti. The analysis presented here was carried out as part of the
Jefferson Lab Hall-B Data-Mining project supported by the U.S. Department of Energy (DOE). The research was supported also by the National Science Foundation, the Pazy Foundation, the Israel Science Foundation, the Chilean Comision Nacional de Investigacion Científica y Tecnologica, the French Centre National de la Recherche Scientifique and Commissariat a l’Energie Atomique, the French-American Cultural Exchange, the Italian Instituto Nazionale di Fisica Nucleare, the National Research Foundation of Korea, and the UK’s Science and Technology Facilities Council. Jefferson Science Associates operates the Thomas Jefferson National Accelerator Facility for the DOE Office of Science, Office of Nuclear Physics under contract DE-AC05-06OR23177. The raw data from this experiment are archived in Jefferson Lab's mass storage silo.
\end{acknowledgments}

\bibliography{TranparencyMeytal_bib}

\end{document}